\documentclass[%
 aip,
rsi,%
 amsmath,amssymb,
 reprint,%
]{revtex4-1}

\usepackage{graphicx}
\usepackage{dcolumn}
\usepackage{bm}

\begin{document}

\preprint{AIP/123-QED}

\title{Slender Origami with Complex 3D Folding Shapes\hfill \break}

\author{Soroush Kamrava}
\affiliation{ 
Department of Mechanical and Industrial Engineering, Northeastern University, Boston, MA 02115, USA
}%
\author{Ranajay Ghosh}%
\affiliation{ 
Department of Mechanical and Aerospace Engineering, University of Central Florida, Orlando, FL 32816, USA
}%
\author{Yu Yang}
\affiliation{ 
Department of Mechanical and Industrial Engineering, Northeastern University, Boston, MA 02115, USA
}%
\author{Ashkan Vaziri}%
 \email{vaziri@coe.neu.edu}
\affiliation{ 
Department of Mechanical and Industrial Engineering, Northeastern University, Boston, MA 02115, USA
}%
\date{\today}

\begin{abstract}
One-dimensional slender bodies can be deformed or shaped into spatially complex curves relatively easily due to their inherent compliance. However, traditional methods of fabricating complex spatial shapes are cumbersome, prone to error accumulation and not amenable to elegant programmability. In this letter, we introduce a one-dimensional origami based on attaching Miura-ori that can fold into various programmed two or three dimensional shapes. We study the out-of-plane displacement characteristics of this origami and demonstrate with examples, design of slender bodies that conform to programmed complex spatial curves. Our study provides a new, accurate, and single actuation solution of shape programmability.
\end{abstract}

\maketitle

The inherent compliance of slender structures makes them easy to deform into complex spatial shapes. This makes them the geometry of choice for a number of biological applications such as DNA scaffolds \cite{han2011dna,hannestad2008self,ke2012three,rothemund2006folding}, microbial appendages \cite{brennen1977fluid,blocker2003type,ghosh2014type} and plant tendrils \cite{goriely1998spontaneous}. Many engineering designs also seek to leverage this deformability in applications such as robotic grippers \cite{amase2015mechanism,voisembert2011novel,nishioka2012proposal,kamrava2018programmable}, deployable structures \cite{puig2010review,hanaor2001evaluation,2018arXiv180710368K}, medical implants \cite{paryab2012uniform,stoeckel2002survey}, prosthetics \cite{weyand2009fastest} and soft robotics \cite{rafsanjani2018kirigami,connolly2017automatic}. However, biological structures still show far greater shape flexibility, functionality and deformation rates, transitioning between multiple shapes over wide time scales from fast protein folding that takes few microseconds \cite{kubelka2004protein} to very slow movements in kingdom plantae \cite{forterre2013slow}. Extracting such wide range of responses has been challenging for man-made structures. However, ability to attain complex geometries is highly desirable since it leads to an expansion of the design space and functionality. Typically one can obtain complex spatial curves either through direct fabrication using conventional manufacturing such as wire draw and metal forming or modern additive manufacturing. These are difficult to adapt for complicated spatial curves due to complexity of the fabrication set up for the conventional process and the complexity of scaffolds, overhangs and sensitivity to process parameters for additive manufacturing.\par
A typical alternative is to start from an easily available thin flat sheet and then crimp it repeatedly to obtain the desired shape. Fig. 1(a) shows a simple example of such a geometry which can be obtained from a straight reference configuration of a slender metallic plate. In this case, multiple crimps (localized bending) were used to shape the straight metallic plate into the desired eight-pointed star like shape. While such localized bending and twisting can be used to form a wide range of shapes, this process, like the ones mentioned earlier, is not reversible due to plastic deformation at the folds and may cause fracture \cite{valiev2002paradox,valiev2004nanostructuring}. In addition, from a fabrication standpoint, this is a multi-step process with multiple sequential crimping operations. This can lead to cumulative addition of deviations from desired shape increasing errors. 
\begin{figure}[t]
\centering
\includegraphics[width=8.7cm]{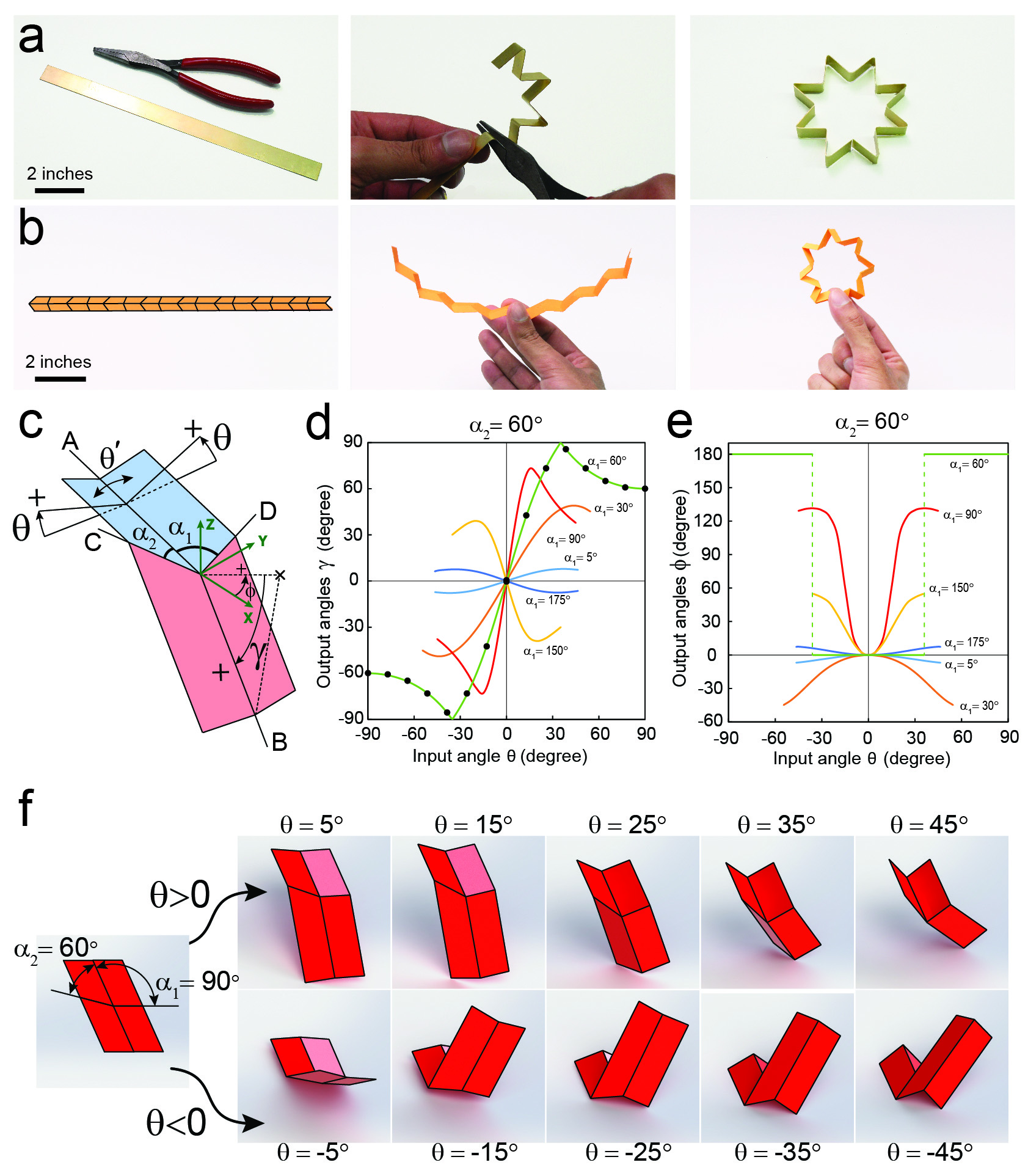}
\caption{\label{fig 1} (a) Shaping a metal strip into an eight-pointed star shape through a sequential crimping. (b)  A one-dimensional origami that evolves into eight-pointed star shape as it folds. (c) In the four-crease pattern, A and B creases are aligned with the longitudinal direction of origami and creases C and D intersect with crease A with angles $\alpha_1$ and $\alpha_2$, respectively. (d,e) Folding response of four-crease pattern. $\gamma$ is the angle between crease B and plane XY and angle $\phi$ represents the out-of-plane angle of origami and is the angle between projection of line B on plane XY and X axis. The markers on $\gamma$ plot show the analytical results for $\alpha_1=\alpha_2=60^{\circ}$. (f) Folding sequence of a four-crease origami with $\alpha_1=90^{\circ}$ and $\alpha_2=60^{\circ}$. The first row shows valley folding $(\theta>0^{\circ})$ and the second row shows mountain folding ($\theta<0^{\circ}$).}
\end{figure}
In contrast, a single-step fabrication technique in which the structure is created by a single actuation event (mechanical, chemical, thermal, etc.) is beneficial in term of accuracy, speed and simplicity. However, a single step crimping would require the use of specific die configurations considerably restricting its generality. On the other hand, shape memory alloys can be programmed into an initial desired shape, which would then be restored through heat \cite{funakubo1987shape} in a single step.  However, programming the shape memory material into complex shapes would require specific molds and ‘chips’ to align with the desired configuration and considerable thermal loads. Both of these processes are therefore, very difficult to scale and adapt for spatial curves. In contrast, folding-based approach such as origami where only the fold is made of actuating/stimuli responsive material can enable a wide range of shapes and patterns using a single actuation event much more conveniently  \cite{overvelde2016three,dudte2016programming}.\par
In this letter, we introduce a one-dimensional slender origami based on attaching Miura-ori \cite{miura1980novel} folds to form a slender body which can fold from a flat reference state into various programmed shapes which could be two or three dimensional as desired. Fig. 1(b) shows an example of such one-dimensional slender origami made out of paper, which evolves into the eight-pointed star based on single folding action. The distinction from the crimping technique is clear in this case, because, the origami, which has one degree of freedom and a single folding action controls global shape.\par
To understand the folding of such slender origami into complex shapes, we study the folding response of the Miura-ori fold shown in Fig. 1(c) in which angles $\alpha_1$ and $\alpha_2$ are not necessarily equal. Angle $\theta$ represents the origami folding angle and varies from $0^{\circ}$ (flat configuration) to the maximum possible value of $90^{\circ}$ (fully folded configuration admissible for a pattern with $\alpha_1=\alpha_2$), see Fig. 1(c). The angle $\theta=\frac{1}{2}(180^{\circ}-\theta^\prime)$, where $\theta^\prime$ is the dihedral angle between two facets sharing a longitudinal crease line (A or B). A fixed right-handed Cartesian coordinate system is attached to the origami structure with origin located on the intersection of crease lines. In our analysis, X axis is aligned with crease line A and Z axis is bisector of the angle $\theta^\prime$. Angle $\gamma$ is measured between XY plane and crease line B and can vary from $-90^{\circ}$ to $90^{\circ}$ during the folding of origami structure with the positive direction convention shown in the figure. The angle $\phi$ is the angle between the X axis and projection of crease B on XY plane. This angle can vary from $-180^{\circ}$ to $180^{\circ}$ with the positive direction convention shown in the figure. The Miura-ori, also known as four-crease pattern, has only one degree of freedom \cite{kamrava2017origami}. Therefore, its configuration at any arbitrary folding level can be fully defined by either $\gamma$, $\phi$ or $\theta$.\par
The relationship between $\theta$ and $\gamma$ or $\theta$ and $\phi$ is highly non-linear and there are no available analytical solutions for them. However, some numerical approaches or restricted analytical solutions for the case of $\alpha_1=\alpha_2$ are available \cite{kamrava2018programmable,hull2002modelling,huffman1976curvature,wu2010modelling}. We simulated the folding of origami in a commercially available software, SolidWorks (Dassault Systems, Vélizy-Villacoublay, France). The simulations solve the rigid body equations of motion for each origami facets numerically to predict geometrically admissible configurations \cite{solidworks2010understanding}.  These simulations estimate the relation between angle $\theta$ and output parameters $\gamma$ and $\phi$ as the origami folds. Fig. 1(d) shows the dependence of $\gamma$ on $\theta$ when $\alpha_2=60^{\circ}$ and $\alpha_1$ varies from $5^{\circ}$ to $175^{\circ}$. $\gamma$ starts from $0^{\circ}$ at $\theta=0^{\circ}$ and as the value of $\theta$ approaches $+90^{\circ}$ or $-90^{\circ}$, $\gamma$ goes back to zero after passing through an extremum point. The folding procedure stops at a folding angle corresponding to contact between facets. This maximum value of $\theta$ can be determined as a function of $\alpha_1$ and $\alpha_2$ (see Supplementary Material for derivation of this equation)
\begin{equation}
 \theta_{max}=\pm(90^{\circ}-\frac{1}{2}\cos^{-1}(\frac{\tan\alpha_2}{\tan\alpha_1})),
\end{equation}
where $|\tan\alpha_1|>|\tan\alpha_2|$. We can compare our numerical simulations with the special case of $\alpha_1=\alpha_2=\alpha$ for which a closed from solution for $\gamma$ as a function of $\theta$ is available in literature \cite{kamrava2018programmable}:
\begin{widetext}
\begin{equation}
\gamma = \left\{ \begin{array}{rcl}
2k\cos^{-1}(\frac{\cos\alpha}{\sqrt{1-\cos^2\theta\sin^2\alpha}})& \mbox{for} & \theta\leq\cos^{-1}(\sqrt{1-\cot^2\alpha})\\2\pi-2k\cos^{-1}(\frac{\cos\alpha}{\sqrt{1-\cos^2\theta\sin^2\alpha}})& \mbox{for} & \theta>\cos^{-1}(\sqrt{1-\cot^2\alpha}),
\end{array}\right.
\end{equation}
\end{widetext}
where $k = 1$ for $\theta\geq0^{\circ}$ and $k = -1$ for $\theta<0^{\circ}$. Note that from Eq. (1), in this case, contact would occur only when $\theta=\pm90^{\circ}$ which is the fully folded state. This is an important practical case for flat folding design. The black markers on the $\alpha_1=\alpha_2=60^{\circ}$ curve, in Fig. 1(d) indicate the analytical results obtained from Eq. (2) which shows an excellent agreement with simulations. Interestingly, Fig. 1(d) also shows that the angle $\gamma$ is an odd function of $\theta$. Physically this indicates that reversing sign of $\theta$ (folding in opposite direction) merely changes the direction but keeps the absolute value of $\gamma$ constant. Fig. 1(e) shows the corresponding variation of angle $\phi$ with $\theta$. This figure shows that as the origami folding proceeds (increasing the absolute value of $\theta$) $\phi$ also increases from zero.  It subsequently achieves an extremum value, which corresponds to the maximum twisting of the Miura-ori and eventually ends at a non-zero value of $\phi$ at $\theta_{max}$ due to contact. In contrast to $\gamma$, $\phi$ is an even function of $\theta$, which means that the direction of folding is immaterial to both the direction and magnitude of $\phi$. These mathematical outcomes are illustrated in Fig. 1(f), which is a set of illustrations from folding of four-crease pattern with $\alpha_1=90^{\circ}$ and $\alpha_2=60^{\circ}$ in both directions ($\pm\theta$). Value of $\gamma$ is positive when it folds downward and negative when it folds upward, and regardless of the folding direction, line B turns toward the larger angle $\alpha$, which satisfies the properties expected of an even function in Fig. 1(e). Four-crease patterns with different $\alpha_1$ and $\alpha_2$ and lengths can be attached together along a straight line to form a slender origami that could fold to a wide range of programmed two dimensional and three dimensional shapes.\par
\begin{figure}[t]
\includegraphics[width=9cm]{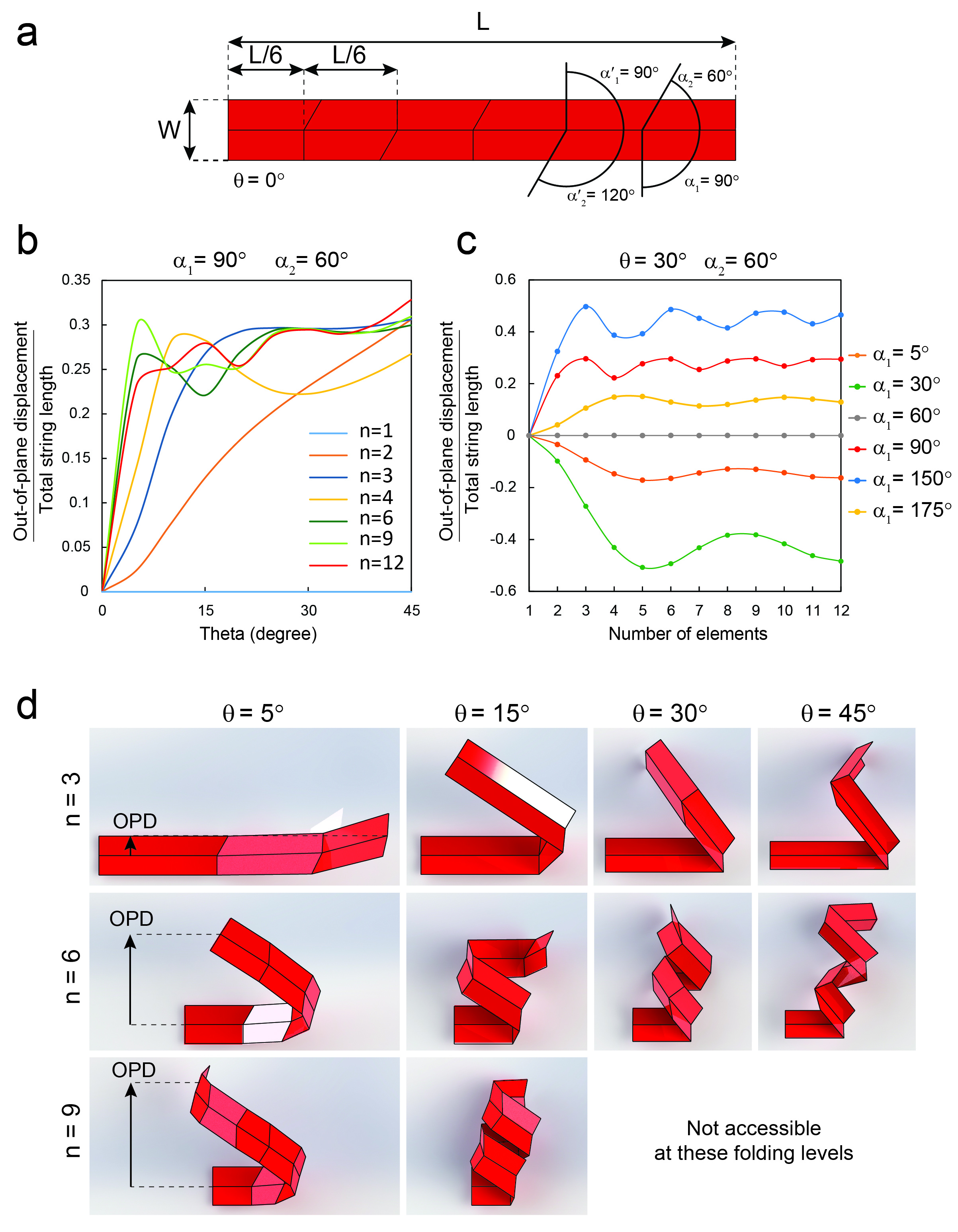}
\caption{\label{fig 2}Out-of-plane displacement of origami string. (a)  Origami string made from five interconnected “four-crease” patterns in a $L\times W$ paper strip with repetitive $\alpha_1=90^{\circ}$, $\alpha_2=60^{\circ}$ and their supplementary angles. (b) Simulation results for the variation of normalized out-of-plane displacement (OPD) as a function of angle $\theta$ for different numbers of segments ($n$). (c) Simulation results for the variation of normalized OPD as a function of $n$ for different angles $\alpha$. (d) Three origami strings with equal lengths ($l$) and 3, 6 and 9 number of segments at four levels of folding ($\theta=5^{\circ}$, $15^{\circ}$, $30^{\circ}$ and $45^{\circ}$). The folded configurations for a string with nine divisions and $\theta>21^{\circ}$ isn't accessible due to the self-intersecting in string.}
\end{figure}
\begin{figure*}[ht]
\centering
\includegraphics[width=18cm]{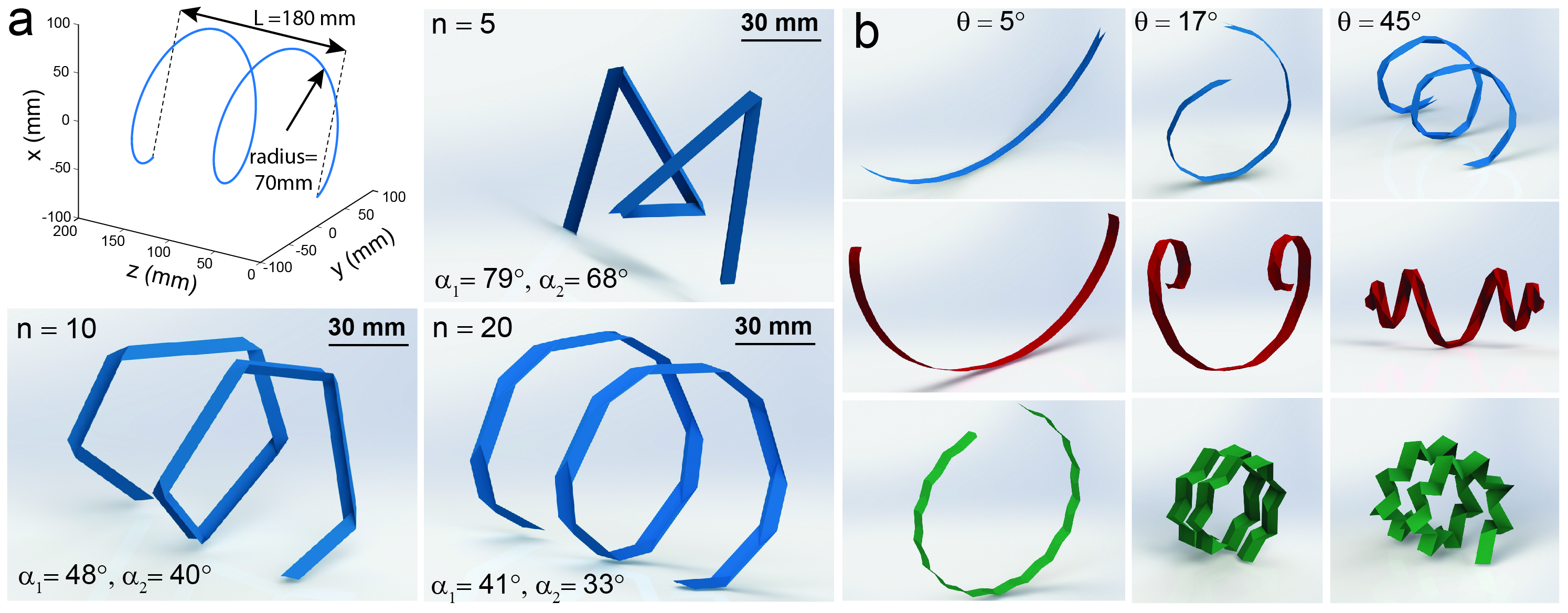}
\caption{\label{fig 3}Designing 3D string. (a) A helix with equation $x(z)^2+y(z)^2=70^2$ is discretized to $n$ (5, 10, and 20) equal segments. The corresponding angles of $\gamma$ and $\phi$ are measured between each two neighboring segments and then values of $\alpha_1$ and $\alpha_2$ are chosen based on the measured values of $\gamma$ and $\phi$. Three origami strings with 5, 10 and 20 segments are shown here which fit to the helix curve with some amount of error and as $n$ increases this error diminishes. (b) Three strips of paper with identical length and width are patterned by different crease line (see Supplementary Material for details on the crease pattern). Three folding levels ($\theta=5^{\circ},17^{\circ}$ and $45^{\circ}$) are shown for each design. Blue, red and green strings fold to a helical, double-spiral and star-helical final shapes at $\theta=45^{\circ}$, respectively.}
\end{figure*}
Next, we study the out-of-plane displacement of a slender origami, which is critical for creating three-dimensional folding shapes because a design with only in-plane displacement would only fold into planar shapes. We consider a slender origami with length L and width W, which comprises of six four-crease patterns with underlying angles $\alpha_1=90^{\circ}$, $\alpha_2=60^{\circ}$, $\alpha^\prime_1=90^{\circ}$ and $\alpha^\prime_2=120^{\circ}$, Fig. 2(a). We quantify this out-of-plane displacement (OPD), as the distance between the free end of the string and XZ plane of the first four-crease pattern in the string. Number of segments along a slender origami is denoted by $n$ and is equal to the number of four-crease patterns plus one. Angles $\alpha_1$, $\alpha_2$, and $n$ are considered as three main characteristics of the presented origami, which can be altered to change the OPD of string. Fig. 2(b) shows the value of $\frac{OPD}{L}$, which is a dimensionless parameter, as a function of $\theta$ (folding level) for different $n$ values, while $\theta$ varies from initial folding angle $0^{\circ}$ to maximum folding angle $45^{\circ}$ determined from Eq. 1. For $n=1$, normalized OPD is always zero by definition but for larger $n$, it always starts from zero and goes to some non-zero value. For $n=2$, normalized OPD increases almost linearly with $\theta$. However, as the number of unit cells increases, such linear and monotonic behavior should not be expected. This is because the position of the free end (tip) of the origami is determined by the complex interaction of rotations of individual units. This would mean that for a given configuration and folding level of an origami, simply increasing the number of units would not necessarily lead to increase in OPD. This is show in Fig. 2(c) where we study the variation of normalized OPD in a folded string ($\theta=30^{\circ}$) with different angles $\alpha_1$ with $\alpha_2=60^{\circ}$ as a function of $n$. The figure shows that for any configuration, origami with more elements correspond to an increasing OPD magnitude for the initial addition of units. However, the increase is not monotonic as the complexity of the origami increases with units. These mathematical insights are summarized pictorially in Fig. 2(d) which illustrates change of OPD during the folding for three strings with $n$ equals to 3, 6 and 9 (angles $\alpha_1$ and $\alpha_2$  are same as Fig. 2(a)) in four levels of folding. The inflections in tip deflections observed in Fig. 2(b,c) can be seen in the changing tip positions with folding level in this figure. The figure also shows that the number of origami units cannot be increased unencumbered since self-contact prevents access to the maximum possible folding range, limiting the design space.\par
Our study so far has shown that using Miura-ori units, the tip of the structure can be raised to a programmed spatial position. However, the real strength of the method comes from an extension of this technique to synthesize more complex spatial curves. We illustrate this by designing an origami, which folds into a helix described by $x=-70\cos⁡(\frac{\pi z}{45}) mm$ and $y=70\sin⁡(\frac{\pi z}{45}) mm$, where $0\leq z\leq 180 mm$, as shown in Fig. 3(a). This is a helix of radius $70 mm$ and pitch $90 mm$. To design the origami to approximate this helix, we divide the helix to $n+1$ equal segments by putting $n$ markers in equal distances along the helix. This means that the coordinates of each marker can be obtained from plugging $z_i=i\frac{L}{n}$ into X and Y expression of the helix equation, where $i=0,1,...,n$ representing the $i^{th}$ marker. Thus the helix is now divided into $n+1$ nodes. Adjacent nodes can be connected by straight line segments leading to $n$ straight line segments. We then treat each pair of adjacent lines as part of a four-crease origami. In this construction, the origin of our coordinate system introduced earlier will be at the intersection of these pair of lines and we will measure the $\gamma$ and $\phi$ angles between them.  The angles $\gamma$, $\phi$ is determined from the geometry of the line segments. Using simulations, we can choose appropriate $\alpha_1$ and $\alpha_2$ which would be the best approximation for $\gamma$ and $\phi$ of a particular four-crease pattern. The angles $\alpha_1$ and $\alpha_2$ do not have to be unique for this design of the helix but would determine the crease pattern along the helix. As the number of line segments approximating the helix increases, the changes in $\gamma$ would be milder giving rise to smoother and better approximations. However, at the same time there is an inherent limit on the number of segments due to self-contact of the origami. Fig. 3(a) shows three designed origami with $n=5,10$, and 20 which mimic the given helix. As expected, when $n$ is increased , the origami better approximates the helix while folding. The values of $\alpha_1$ and $\alpha_2$ repeating along the entire slender origami are shown in the bottom left corner of each picture.\par
The same procedure can be implemented to design origami, which fold to other more complex shapes from a flat reference state which can serve applications such as robotic manipulator \cite{kamrava2018programmable,edmondson2013oriceps}, deployable space structures \cite{zirbel2013accommodating} and foldable building blocks \cite{mousanezhad2017origami}. In Fig. 3(b), we illustrate the folding procedure of three examples including helix, double-spiral, and star-shape helix with identical unfolded shape and completely different folded configurations. See Supplementary Material for more details about these three designs.\par
In conclusion, our work provides an alternative to design 3D space curves out of a flat and thin sheet to other techniques such as discretized rigid-foldable curvatures \cite{demaine2011curved}, continues buckled curvature \cite{dias2012geometric,dias2012shape} and tessellated origami patterns to approximate a 3D geometry \cite{dudte2016programming}. However, this method is distinct in providing a simple way to fabricate spatial shapes using a single actuation regardless of the complexity of desired pattern. This technique overcomes many of the limitations of traditional fabrication techniques.

\begin{acknowledgments}
This work is supported by the United States National Science Foundation, Division of Civil, Mechanical, and Manufacturing Innovation, Grant No.1634560.
\end{acknowledgments}
\nocite{*}
\bibliography{aipsamp}

\end{document}